\documentclass[usenatbib]{mn2e}
\usepackage{graphicx}
\usepackage[english]{babel}
\title[Spectroscopy of `SN 2003aw' in quiescence]{Phase-resolved spectroscopy of the helium dwarf nova `SN 2003aw' in quiescence}
\author[G.\,H.\,A. Roelofs et al.]{G.\,H.\,A.~Roelofs,$^1$\footnote{E-mail: g.roelofs@astro.ru.nl} P.\,J.~Groot,$^1$ T.\,R.~Marsh,$^2$ D.~Steeghs$^3$ and G.~Nelemans$^1$\\
$^1$Department of Astrophysics, Radboud University, PO Box 9010, 6500 GL Nijmegen, The Netherlands\\
$^2$Department of Physics, University of Warwick, Coventry CV4 7AL, UK\\
$^3$Harvard-Smithsonian Center for Astrophysics, 60 Garden Street, Cambridge, MA 02318, USA}

\newcommand{\obj}{`SN\,2003aw'}
\begin{document}
\maketitle

\begin{abstract}
High time resolution spectroscopic observations of the ultra-compact helium dwarf nova \obj\ in its quiescent state at $V\!\sim\!20.5$ reveal its orbital period at $2027.8\!\pm\!0.5$ seconds or 33.80 minutes. Together with the photometric ``superhump'' period of $2041.5\!\pm\!0.5$ seconds, this implies a mass ratio $q\!\approx\!0.036$. We compare both the average and time-resolved spectra of \obj\ and SDSS J124058.03$-$015919.2 \citep{roelofs}. Both show a DB white dwarf spectrum plus an optically thin, helium-dominated accretion disc. \obj\ distinguishes itself from the SDSS source by its strong calcium H \& K emission lines, suggesting higher abundances of heavy metals than the SDSS source. The silicon and iron emission lines observed in the SDSS source are about twice as strong in \obj. The peculiar ``double bright spot'' accretion disc feature seen in the SDSS source is also present in time-resolved spectra of \obj, albeit much weaker.
\end{abstract}


\section{Introduction}

The AM CVn stars are ultra-compact binaries: white dwarfs accreting from a degenerate, helium-rich companion star. They have orbital periods shorter than about one hour, clearly indicating the evolved nature of their donor stars. Their accretion disc spectra show mainly helium, with no traces of hydrogen and varying contributions of heavy metals. See \citet{nelemans} for a recent review.

No less than six new members of the small AM CVn family, including the first eclipsing system, have been found in the last year or so \citep{wwnew,roelofs,anderson}. Optical spectroscopy of one of the new members, SDSS J124058.02$-$015919.2 (hereafter SDSS J1240, \citealt{roelofs}) allowed for a measurement of the orbital period and of the mass ratio, and showed a peculiar and as yet not understood feature of two equally strong bright spots in the accretion disc, in strong contrast with the usual one bright spot attributed to the impact of the accretion stream into the disc. In this paper we present time-resolved spectroscopy of another recent addition to the family, and a close relative of SDSS J1240: \obj.

\obj\ was found as a supernova candidate in February 2003, based on an observed strong increase in brightness (maximum $V\!\sim\!16.5$) and its supposed positional coincidence with a faint galaxy \citep{woodvasey}, but \citet{chornock} identified it spectroscopically as a helium-rich dwarf nova. Following their announcement, \citet{ww03} performed fast white-light photometry on the source, from which a periodic modulation of $2041.5\pm0.5$ seconds was derived. They attributed this period to a superhump, although an analysis of the Fourier spectrum showed sidebands to the main period that could point to an orbital modulation rather than a superhump.

Over a period of about six months from its detection as a supernova candidate the source was seen to decline to its apparent quiescent state at $V\!\sim\!20.5$ \citep{ww03}. We caught the star in quiescence in December 2003, when we took a few low-resolution spectra. Our first set of phase-resolved spectroscopic observations was taken in March 2004, thirteen months after its discovery, with the star still in quiescence near $V\!=\!20.5$. In May 2004 the star was reported to be in outburst again \citep{nogami} with an even higher maximum detected brightness at $V\!=\!15$. In March 2005 we then obtained our second, larger set of phase-resolved spectra, the star having returned to a quiescent state again near $V=20.5$.

\section{Observations and data reduction}

\begin{table}
\begin{center}
\begin{tabular}{l l r r}
\hline
Date		&UT		        &Exposure       &Exposures\\
                &                       &time (s)       &\\
\hline
\hline

Magellan&&&\\

2003/12/15      &08:08--08:40           &900            &3\\
&&&\\

VLT&&&\\

2004/03/18	&01:27--02:43	        &260	        &14\\
2004/03/19	&00:24--01:23	        &260	        &14\\
2005/03/01	&00:30--05:04	        &180	        &70\\
2005/03/02	&00:37--03:33	        &180	        &39\\
        	&03:38--05:22	        &240	        &22\\

\hline
\end{tabular}
\caption{Summary of our observations of \obj.}
\label{observations}
\end{center}
\end{table}

We obtained phase-resolved spectroscopy of \obj\ on 18 and 19 March 2004, and on 1 and 2 March 2005 with the Very Large Telescope (VLT) of the European Southern Observatory and the FOcal Reducer/low dispersion Spectrograph (FORS2). The observations consist of 159 spectra in the blue (grism 600B), most of them with a 180-second exposure time. Typical seeing was $0.6''$ giving an effective resolution of $\sim$4\,\AA; for 50 of the 159 spectra we used slightly longer exposure times of 240 or 260 seconds because of less perfect seeing conditions ($\sim$$1''$, resolution $\sim$6\,\AA).

All VLT observations were done with a $1''$ slit. The detector was the MIT CCD mosaic of which only chip 1 was used; binning was standard $2\times2$ pixels. Low read-out speed and high gain minimized the read-out and digitisation noise, respectively. In order not to add more noise to the spectra, we subtracted a constant bias level from each CCD frame. The bias level was determined per frame from the overscan regions on the CCD. A normalised flatfield frame was constructed from 5 incandescent lamp flatfield frames each night, which ensured a cosmic-ray-free final flatfield.

All spectra were extracted using the IRAF implementation of optimal (variance-weighted) extraction. The read-out noise and photon gain, necessary for the extraction, were calculated from the bias and flatfield frames, respectively. Wavelength calibration was done with a standard HeHgCd arc exposure taken during the day. A total of around 40 arc lines could be fitted well with a Legendre polynomial of order 3 and 0.14\,\AA\ root-mean-square residual. All spectra were transformed to the heliocentric rest-frame prior to analysis.

The average spectrum was corrected for instrumental response using spectroscopic standard star EG274.

In addition to the phase-resolved spectra described above, we obtained three 900-second spectra with the IMACS spectrograph on the Magellan--Baade telescope at Las Campanas Observatory on 15 December 2003. An $0.7''$ slit together with the 300 lines/mm grating provided coverage between 3400--9500 angstrom at 4.6\,\AA\ resolution.

A summary of all observations is given in table \ref{observations}.

\section{Results}

\subsection{Average spectrum}
\label{averagespectrum}

\begin{figure*}
\centering
\includegraphics[angle=270,width=\textwidth]{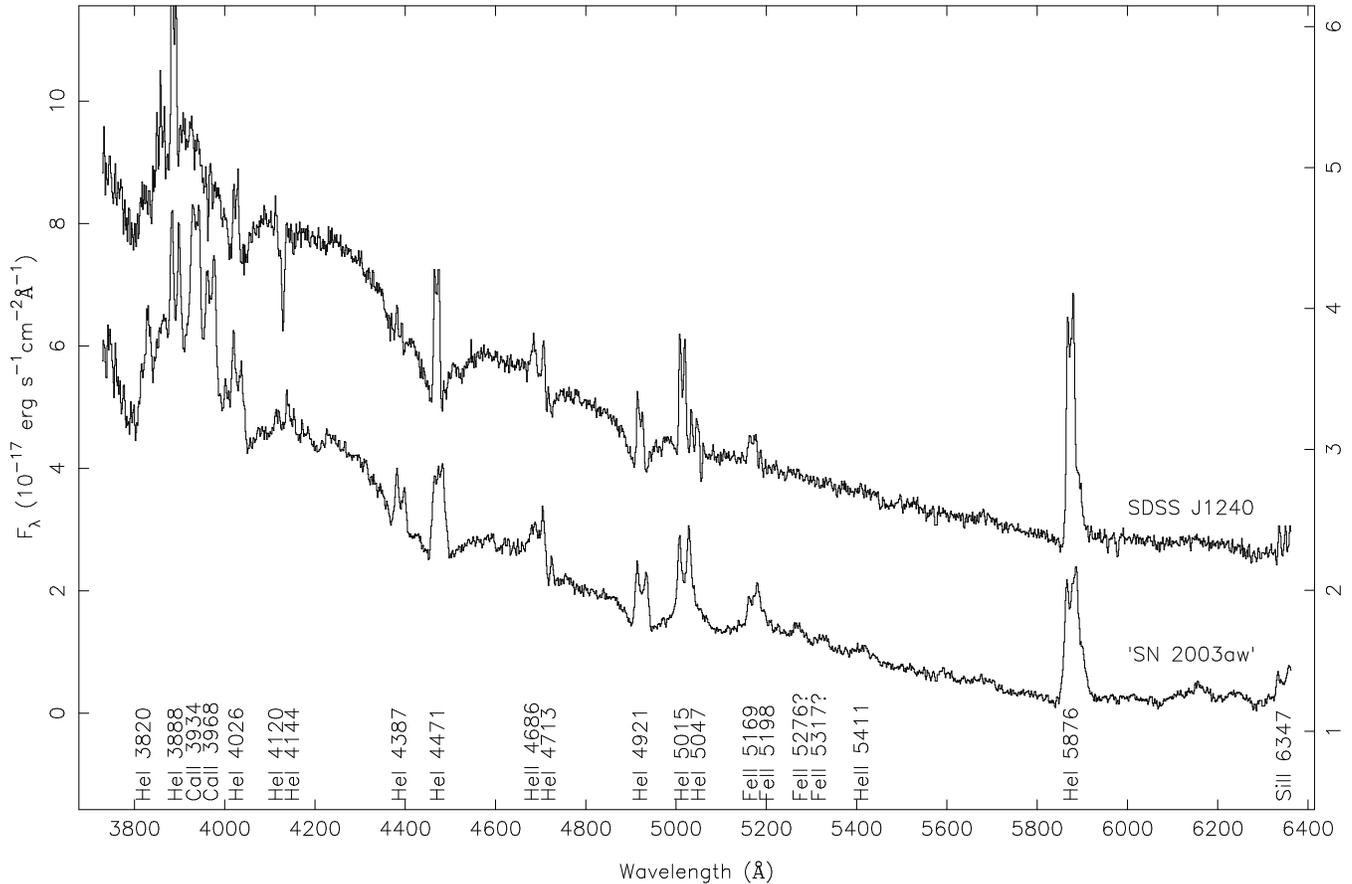}
\caption{Average spectra of SDSS J1240 (top spectrum \& left axis, \citealt{roelofs}) and \obj\ (bottom spectrum \& right axis). The most prominent lines are labelled. Uncertain identifications carry a question mark.}
\label{average}
\end{figure*}

The flux-calibrated average spectrum of \obj\ is shown in figure \ref{average}. It is a striking match to the average blue spectrum of SDSS J1240 from \citet{roelofs}. The latter is reproduced in figure~\ref{average} for easy reference. The spectrum of \obj\ also shows the characteristic broad helium absorption lines, presumably from the accreting white dwarf, in addition to the double-peaked emission lines, mainly helium, from the accretion disc. There are a few differences: first, the accretion disc lines are much broader, extending to $\sim$1000 km/s FWHM, suggesting a higher inclination than SDSS J1240. Second, the accretion disc shows strong calcium H \& K emission, in addition to the iron and silicon features also seen in SDSS J1240. Table \ref{eqw} lists the equivalent widths of a number of accretion disc emission lines for both SDSS J1240 and \obj; the blue helium lines are omitted because of the strong influence of the underlying white dwarf's absorption lines. The strengths of the helium lines are identical in both objects, whereas the heavy metal lines are significantly stronger in \obj. The \mbox{He\,{\sc i}} 5876 line is blended with a weak unidentified emission feature around 5896\,\AA\ in both objects, which might be Na D. This emission feature is included in the equivalent widths of the \mbox{He\,{\sc i}} 5876 line in table \ref{eqw}.

\begin{table}
\begin{center}
\begin{tabular}{l r r}
\hline
Line		                &\multicolumn{2}{c}{Equivalent widths (\AA)}\\
                                &SDSS J1240$^{1}$             &\obj$^{2}$\\
\hline
\hline

\mbox{He\,{\sc i}} 5876         &$-31.3\pm 0.5$               &$-30.3\pm 0.5$\\
\mbox{He\,{\sc i}} 6678         &$-19.6\pm 0.1$               &$-18.5\pm 1.1$\\
\mbox{He\,{\sc i}} 7065         &$-25.0\pm 0.1$               &$-25.5\pm 1.7$\\
\mbox{He\,{\sc i}} 7281         &$-10.3\pm 0.1$               &$-11.3\pm 2.2$\\
\mbox{Fe\,{\sc ii}} 5169        &$-2.1\pm 0.2$                &$-5.0\pm 0.5$\\
\mbox{Si\,{\sc ii}} 6347        &$-2.2\pm 0.1$                &$-4.0\pm 0.4$\\
\mbox{Ca\,{\sc ii}} 3934        &$0.0\pm 0.5$                &$-5.6\pm 0.5$\\
\mbox{Ca\,{\sc ii}} 3968        &$0.0\pm 0.5$                &$-5.6\pm 0.5$\\

\hline
\end{tabular}
\caption{Equivalent widths of several accretion disc emission lines, including estimated errors. $^{1}$Spectra published in \citet{roelofs}; $^{2}$\mbox{He\,{\sc i}} 6678, 7065 \& 7281 lines based on our Magellan spectrum (not shown).}
\label{eqw}
\end{center}
\end{table}

\subsection{The spectroscopic period}
\label{spectroscopicperiod}

\begin{figure}
\includegraphics[angle=270,width=84mm]{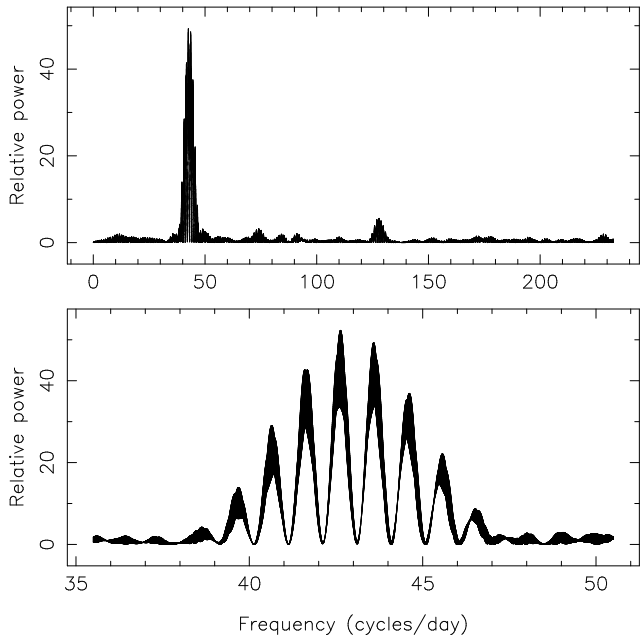}
\caption{Lomb-Scargle periodogram of the red wing/blue wing emission line flux ratios. The lower panel provides a magnified view of the strongest peaks.}
\label{scargle}
\end{figure}

To determine the spectroscopic period of the binary, we used a modified version of the method used by \citet{nather}, as described in \citet{roelofs}. Figure \ref{scargle} shows the resulting Lomb--Scargle periodogram of red wing--blue wing emission line flux ratios. The long baseline between our 2004 and our 2005 observations causes a fine pattern of aliases on top of the usual 1 day$^{-1}$ aliases in the periodogram. The strongest peak occurs at 42.6 cycle/day. In addition to the main peaks around this frequency there appears to be a weak group of higher harmonics much like in SDSS J1240 \citep{roelofs}. The strongest harmonic is exactly three times the frequency of the overall strongest peak, compatible with a second bright spot in the accretion disc appearing 120 degrees out of phase.

\subsection{Doppler tomography}
\label{doppler}

\begin{figure*}
\centering
\includegraphics[angle=270,width=\textwidth]{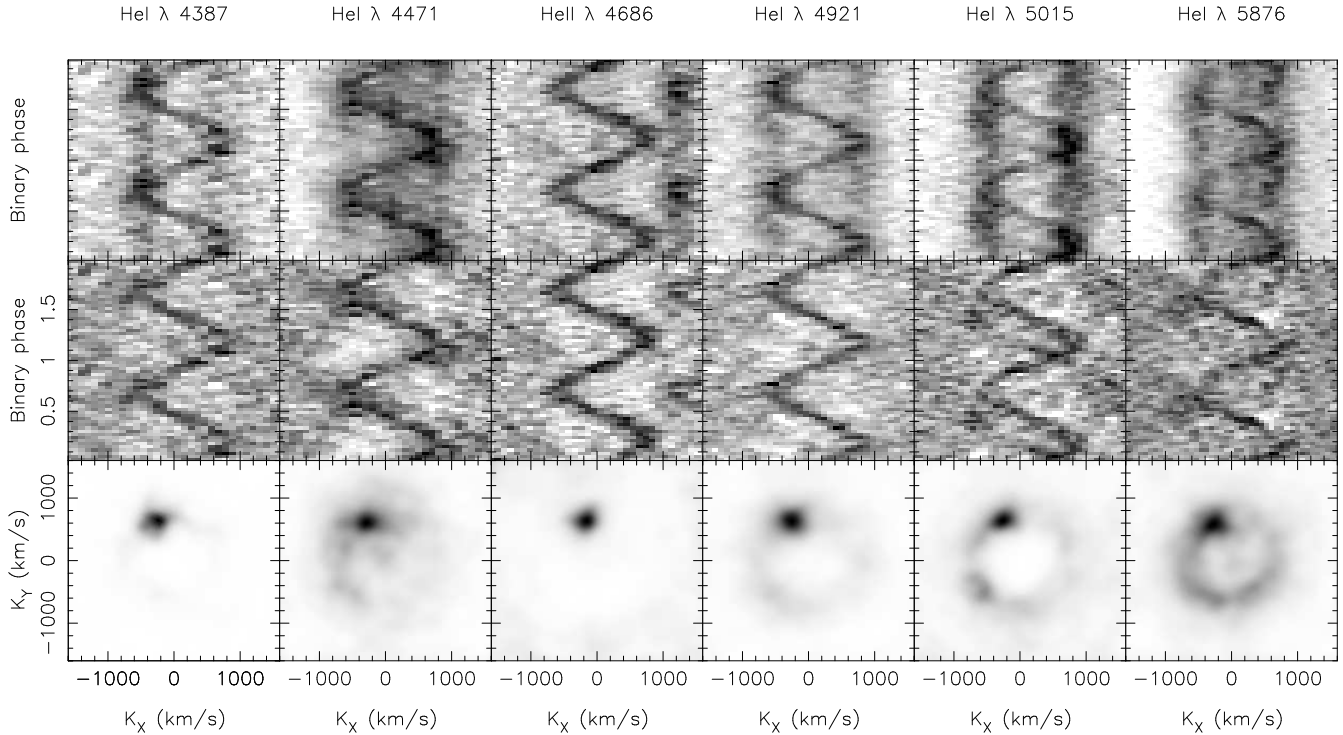}
\caption{Trailed spectra (top row), average-subtracted trailed spectra (middle row) and maximum-entropy Doppler tomograms (bottom row) of the strongest \mbox{He\,{\sc i}} and \mbox{He\,{\sc ii}} features.}
\label{tomograms}
\end{figure*}

\begin{figure}
\includegraphics[angle=270,width=84mm]{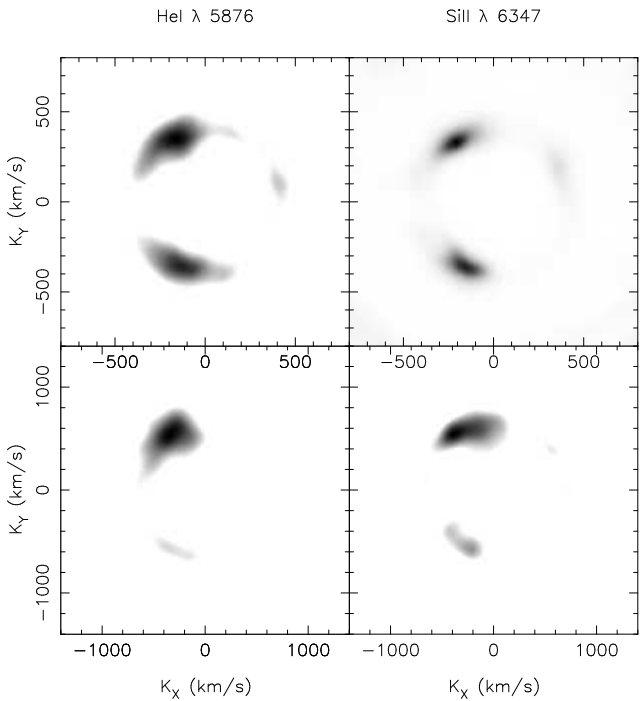}
\caption{Detailed comparison of Doppler tomograms between SDSS J1240 (top row) and \obj\ (bottom row). The tomograms of \obj\ are aligned with those of SDSS J1240 artificially, since we do not have a zero-phase measurement for \obj. A linear back-projection code that can cope with the \mbox{Si\,{\sc ii}} line being at the edge of our spectral window is used for \obj.}
\label{tomogramscompared}
\end{figure}

\subsubsection{Features of the accretion disc}

Figure \ref{tomograms} shows the trailed spectra and maximum-entropy Doppler tomograms \citep{dopplermapping} of the strongest lines of \obj. There is one strong emission bright spot causing a clear S-wave in the trailed spectra; the \mbox{He\,{\sc i}} 5015 line also shows a second S-wave at approximately 30\% of the flux of the main S-wave. There are two lines in the current data that are also present in the SDSS J1240 data of \citet{roelofs}, namely \mbox{Si\,{\sc ii}} 6347 and \mbox{He\,{\sc i}} 5876; these are compared directly in figure \ref{tomogramscompared}. The double bright spot pattern in the SDSS source is not reproduced in \obj, but there do seem to be very weak secondary bright spots at approximately the same position as in the SDSS source.

\subsubsection{The orbital period}
\label{orbitalperiod}

In section \ref{spectroscopicperiod} we determined the spectroscopic period from a periodogram of red wing--blue wing emission line flux ratios. The clear kinematic S-wave feature in the trailed spectra, when phase-folded on this spectroscopic period, indicates that we are seeing the orbital motion of the binary. The strength and sharpness of the bright spot in the Doppler tomograms, especially in the \mbox{He\,{\sc ii}} 4686 line, allows us to determine the orbital period quite accurately. The coverage per night during our 2005 run (about 4.5 hours) is sufficient that we can distinguish between the true orbital period and its 1 day$^{-1}$ aliases in the periodogram, since these aliases already lead to noticeable smearing of the bright spot signal when phase-folding the spectra from a single night.

Assuming that the bright spot is fixed in the binary frame, we can further refine the orbital period by lining up the phases of the bright spot in the tomograms of March 1 and March 2, 2005. The bright spot phases are determined by fitting a 2-D Gaussian in the Doppler tomograms. We estimate the uncertainty in the fitted bright spot phases with a simple Monte Carlo simulation, where we make a large ensemble of Doppler tomograms from our dataset using the bootstrap method. For each March 1 Doppler tomogram we thus randomly pick 70 spectra out of the set of 70 spectra that we have for this night, allowing for a spectrum to be picked more than once. We fit a 2-D Gaussian to each tomogram in the ensemble; the resulting distribution of fitted bright spot locations in $K_X, K_Y$ space gives an estimate of the accuracy allowed by the data.

In practice there may also be an intrinsic shift in the bright spot phase due to changes in the effective accretion disc radius at the stream--disc impact point. In such a scenario, the radial velocity of the bright spot in the Doppler tomograms is expected to change more than its phase, since both the accretion stream and accretion disc velocity (at the stream--disc impact point) change faster with accretion disc radius than their directions, for reasonable accretion disc radii (see, for instance, figure 5 in \citealt{roelofs}). The radial velocity change measured in the bright spot between March 1 and 2, 2005 thus gives us a reasonable measure for the maximum intrinsic bright spot phase shift. We add this radial velocity change to the statistical fit uncertainty from the Monte Carlo ensemble; both give an uncertainty of about 20 km/s. The total uncertainty in lining up the bright spots from March 1 and 2, 2005 thus comes out at $\sim$4 degrees. The orbital period and its error then become $P_\mathrm{orb}=2027.8\pm0.5$ seconds.

The one-year baseline between our 2004 and 2005 runs does not allow us to trace the orbit all the way in between, since one revolution fewer or more between these runs would correspond to a phase drift of just about one degree between the two consecutive nights of our 2005 run, which is below the accuracy with which we can measure the bright spot's phases in the 2005 data.

\section{Discussion}

\subsection{Chemical composition of the accretion disc}

\citet{trm91} first modelled the accretion disc spectrum of the 46-minute orbital period AM CVn star GP Com and found a strong underabundance of heavy metals relative to the Sun. In particular, expected emission lines of singly ionised iron, silicon and calcium were missing from the spectrum. The recently discovered AM CVn star SDSS J1240 showed clear \mbox{Fe\,{\sc ii}} 5169, \mbox{Si\,{\sc ii}} 6347 and \mbox{Si\,{\sc ii}} 6371 emission, compatible with solar abundances \citep{roelofs}. Calcium H \& K were also expected if solar abundances of heavy metals were assumed, but could not be found in the spectrum. They do show up strongly in the spectrum of \obj\ presented here. The spectrum of \obj\ furthermore shows possible weak iron features near 5276 and 5317 \AA\ not seen in SDSS J1240 (cf.\ figure \ref{average}). The possible Na D emission feature in both SDSS J1240 and \obj, mentioned in section \ref{averagespectrum}, must originate in relatively cool parts of the accretion disc --- a large part of the disc where most of the emission lines originate is expected to be at $\approx$11,000\,K (see \citealt{trm91} for a discussion), where the sodium should be largely ionised and the Na D lines would not show up. The feature does not show any bright spot (nor the typical double-peaked profile) in the trailed spectra, which fits with being Na D or a similar low-temperature feature.

We now have two emission-line systems that seem to be strongly underabundant in heavy metals (GP Com, $P_\mathrm{orb}=46$\,min and V396 Hya (=CE~315), $P_\mathrm{orb}=65$\,min), which together with their high proper motions suggests a halo origin, and two systems showing more or less the expected metal lines if solar abundances of heavy metals are assumed (SDSS J1240, $P_\mathrm{orb}=37$\,min and \obj, $P_\mathrm{orb}=34$\,min). The stronger silicon and iron lines in \obj\ relative to helium (by about a factor of 2, see table \ref{eqw}) can probably be accommodated by a higher column density of gas in the disc -- e.g.\ due to a higher mass transfer rate -- and saturated helium lines. The complete absence of calcium H \& K in SDSS J1240 is more difficult to explain, but together with the notion that the \mbox{Si\,{\sc ii}} 6347 \& 6371 lines in SDSS J1240 originate exclusively from the bright spots, this suggests somewhat higher abundances of heavy metals in \obj\ (if we assume both discs to be largely at $\approx$11,000\,K). A more detailed study of these optically thin helium-dominated accretion discs is beyond the scope of this paper, but will be interesting for estimating the mass transfer rates and for putting more accurate constraints on the chemical abundances in the discs.

\subsection{The second bright spot}

The remarkable double bright spot feature in SDSS J1240 (see \citet{roelofs} for a discussion on its possible origins) was one of the motivations for obtaining the observations presented here. Since the average spectrum of \obj\ looked so similar, and since the suggested orbital period from the photometric superhump was quite close to that of SDSS J1240, it was interesting to see if the accretion disc would show the same behaviour.

From the trailed spectra and Doppler tomograms (figure \ref{tomograms}) it is clear that the \mbox{He\,{\sc i}} 5015 line does indeed show a second S-wave much like in SDSS J1240, but it is significantly weaker at about 30\% of the integrated flux of the main S-wave. A direct comparison of the \mbox{He\,{\sc i}} 5876 and of the \mbox{Si\,{\sc ii}} 6347 line, for which there are phase-resolved spectra of both objects, shows the weakness of the second bright spot in these lines compared to SDSS J1240 (figure \ref{tomogramscompared}). So more than anything else, \obj\ adds to the diversity of the second bright spot feature. A very recently discovered new AM CVn star, 2QZ J142701.6$-$012310 \citep{wwnew} with a suggested orbital period near 36 minutes (again based on a photometric superhump period), even closer to the orbital period of SDSS J1240, may be a nice test case for this still enigmatic feature.

\subsection{The orbital period}

We measure an orbital period $P_\mathrm{orb}=2027.8\pm0.5$ seconds. This proves that the photometric period of $2041.5\pm0.5$ seconds first detected by \citet{ww03} is a superhump period. Therefore the helium dwarf nova \obj, like the permanent superhumper AM CVn \citep{nsg}, exhibits positive superhumps, which are usually explained as the beat period of the orbit and the prograde precession of a tidally deformed eccentric accretion disc. If we employ the latest empirical relation between the mass ratio $q$ and the superhump period excess $\epsilon$, as determined from superhump and orbital periods of a large number of hydrogen-rich dwarf novae by \citet{patterson}:
\begin{equation}
\epsilon \left(q\right) = 0.18q + 0.29q^2
\label{superhump}
\end{equation}
where
\begin{equation}
\epsilon \equiv \frac{P_\mathrm{sh} - P_\mathrm{orb}}{P_\mathrm{orb}}
\end{equation}
we find $q=0.036\pm0.003$. This compares to $q=0.039\pm0.010$ determined kinematically for SDSS J1240 \citep{roelofs}. It should be stressed that the relation (\ref{superhump}) may not be well-calibrated for these extreme mass ratios, and in reality $\epsilon$ might depend on more parameters than just $q$. The quoted error on the mass ratio may therefore be a bit optimistic.

\section{Acknowledgments}

GHAR and PJG are supported by NWO VIDI grant 639.042.201 to P.J. Groot. DS acknowledges a Smithsonian Astrophysical Observatory Clay Fellowship. GN is supported by NWO VENI grant 639.041.405 to G. Nelemans. TRM was supported by a PPARC Senior Research Fellowship. We thank Patrick Woudt at the University of Cape Town for kindly observing \obj\ for us in preparation of our observing run. This work is based on data taken at the European Southern Observatory, Chile, under programmes 074.D-0662(A) and 072.D-0052(A).

\end{document}